\pdfoutput=1

\documentclass[reprint]{elsarticle}

\usepackage{graphicx}
\usepackage{amsmath}
\usepackage{amssymb}
\usepackage{mathrsfs}
\def\be{\begin{equation}}
\def\ee{\end{equation}}
\def\ba{\begin{array}}
\def\ea{\end{array}}
\def\bea{\begin{eqnarray}}
\def\eea{\end{eqnarray}}

\usepackage{lineno}
\journal{Journal of \LaTeX\ Templates}

\begin{document}
\begin{frontmatter}

\title{Controlled self-similar matter waves in PT-symmetric waveguide}

\author[mymainaddress]{Shailza Pathania}
%%\ead[url]{shalu.pathania7@gmail.com}

\author[mymainaddress,mysecondaryaddress]{Harneet Kaur}
%%\ead[url]{harneet.kaur13@gmail.com}

\author[mysecondaryaddress2]{Amit Goyal\corref{mycorrespondingauthor}}
\cortext[mycorrespondingauthor]{Corresponding author}
\ead{amit.goyal@ggdsd.ac.in}

\author[mymainaddress]{C. N. Kumar}
%%\cortext[mycorrespondingauthor]{Corresponding author}
\ead{cnkumar@pu.ac.in}

\address[mymainaddress]{Department of Physics, Panjab University Chandigarh 160014, India}
\address[mysecondaryaddress]{Department of Physics, Government College for Women, Karnal 132001, India}
\address[mysecondaryaddress2]{Department of Physics, GGDSD College, Chandigarh 160030, India}

\begin{abstract}
 We study the dynamics of Bose-Einstein condensate coupled to a
 waveguide with parity-time symmetric potential in the
 presence of quadratic-cubic nonlinearity modelled by
 Gross-Pitaevskii equation with external source. We employ the
 self-similar technique to obtain matter wave solutions, such as
 bright, kink-type, rational dark and Lorentzian-type self-similar waves for
 this model. The dynamical behavior of self-similar matter waves
 can be controlled through variation of trapping potential,
 external source and nature of nonlinearities present in the
 system.

\end{abstract}

\begin{keyword}
 Matter waves \sep Self-similar solutions \sep PT-symmetry \sep Gross-Pitaevskii equation
\MSC[2010] 35C08\sep 35Q60 \sep 78A60
\end{keyword}

\end{frontmatter}

%\linenumbers

\section{Introduction}

Bose-Einstein condensate (BEC) is a macroscopic quantum state of
matter in which all atoms in
   the bosonic gas condenses into a
 single ground state of the system \cite{Anderson,Davis}. The coherent matter-wave formed
 via this population of atoms is depicted by a macroscopic
 wave function  which is also a solution of nonlinear
Schr\"odinger equation (NLSE). The mean-field equation used to
describe the dynamics of BEC is the
Gross-Pitaevskii (GP) equation or NLSE with a trapping potential
\cite{Pitaevskii}. Basically, the GP equation is a three-dimensional equation which can be reduced to a one- or two-dimensional equation by confining the condensate to two or one directions in an effective potential \cite{nicola,wang}.
 In BECs, the formation of soliton solutions is resultant of the interactions
  among atoms and the geometry of the trap used to confine the BEC. Here, nonlinearity arises due to interatomic
interactions of the condensate measured by the scattering length
`$a$'. In BEC, Feshbach resonance gives a way to control the
strength of how atoms interact with each other \cite{Andrews}.
Experimental work has been done to show the existence of bright
solitons for attractive interactions ($a<0$) \cite{Strecker} and dark solitons for repulsive interactions ($a>0$) \cite{Burger}.
Authors have also studied the dynamics of BEC in the presence of competing
cubic-quintic nonlinearity \cite{Tang,beitia,Avelar,cardoso2,cardoso3,Carlos,kha} and
quadratic-cubic nonlinearity \cite{chaos,cardoso4,optik,pal,Ritu}. Over the past several years,
there is a considerable interest on the existence of matter wave solutions
for GP equation with time-dependent coefficients or generalized nonlinear Schr\"odinger equation (GNLSE) \cite{kevre,PRL,PRE,Avelar2,kumar}.
Earlier, Paul and his collaborators \cite{Paul,Paul2,Paul3}
numerically studied the resonant transport of interacting BEC
through a symmetric double barrier potential in a waveguide for
the modified GP equation. Coupling of the waveguide to a reservoir
of condensate from which matter waves are injected into the guide
is modelled by source term. Later, Yan et al. \cite{Yan} studied
the nonautonomous matter waves in a waveguide for the modified GP
equation driven by a source term.
Recently, R. Pal et al. \cite{Ritu} obtained the matter wave
self-similar solutions for the driven nonautonomous GP equation
with quadratic-cubic nonlinearity. Apart from  it, the GNLSE with
external source has also been studied to obtain self-similar
solution in the context of fiber optics \cite{Raju1,Raju2,he}.

\par In this work, we have studied the dynamics of BEC coupled to
a waveguide with parity-time (PT) symmetric potential modelled by
GP equation with inhomogeneous source, $S(t)e^{i \theta(x,t)}$ where $S(t)$ and $\theta(x,t)$
are amplitude and phase terms \cite{Paul2,Yan}. The source term simulates the coherent
injection of matter waves from an external reservoir to the waveguide.
In recent years, a significant work has been done on the evolution of
soliton \cite{PT2008,Khare,Wen,chen} and self-similar solutions \cite{Dai,
Wang,Deng} for the GNLSE with PT-symmetric potential. According to
quantum mechanics, Hamiltonian of a system should be Hermitian
since the eigenvalues corresponding to Hermitian operators are
always real. In 1998, Carl Bender and S. Boettecher \cite{bender}
proposed that even non-Hermitian Hamiltonians exhibit real spectra
provided Hamiltonians respect PT-symmetry. The necessary but not
sufficient condition for Hamiltonian to be PT-symmetric is that
the real and imaginary parts of the potential should be even and
odd function w.r.t coordinates, respectively \cite{bender2}.
Initially, the idea of PT-symmetry was introduced in the field of
quantum mechanics \cite{bender} and then this idea has found rapid
applications in numerous other fields. In the field of optics, PT
was implemented by Christodouldes and his collaborators
\cite{gana,mussli2008,makris2008} by choosing the complex
potential as PT-symmetric such that the real part models the
waveguide profile and imaginary part models gain/loss in the
media. In refs. \cite{Achilleos,Car,yu}, authors have studied the
nonlinear model for the dynamics of BEC in PT-symmetric potential.
In the context of BEC arrays, work has been done to study the nonlinear excitations in the presence of uniform distribution of atomic population which refer to as uniform background \cite{back1}. The matter wave solutions with background is explored in different trap geometries and interactions \cite{back1,back2,back3,back4}. Motivated from the above works, we consider driven
quadratic-cubic GP equation in the presence of PT-symmetric
potential and report the existence of bright, kink, rational dark and Lorentzian-type self-similar matter wave solutions on uniform background for
this model. Kink solitons (also known as domain walls) have been reported for two-component condensates with cubic nonlinearity \cite{kink1,kink2}. These domain walls represent a transient layer between semi-infinite domains carrying different components, or distinct combinations of the components. Later on, the kink solutions also obtained for single component condensates with cubic-quintic nonlinearity \cite{kink3,kink4}. In earlier works, our group has done a considerable work
on the self-similar solutions for GNLSE and proposed an analytical approach to
control the dynamics of these solutions \cite{kumar,pra,de}. Here, in this work, we control the
dynamical behavior of self-similar matter waves by varying the trapping potential,
nonlinearity coefficient and source profile.
\par The manuscript is organized as follows : In Section 2, we discuss
about the model equation and self-similar technique. In section 3,
we describe the self-similar matter wave solutions for different
profiles of trapping potential. Section 4 summarizes the work.

\section{Model equation}
The dynamics of BEC coupled to a waveguide in the presence of
quadratic-cubic nonlinearity modelled by modified GP equation with external
source as \be\label{e1} i \hslash
\frac{\partial\psi}{\partial t}=\Bigg(-\frac {\hslash^2}{2 m}
\frac{\partial^2}{\partial x^2}+ V(x,t)+ g_{1D}(t) |\psi|^2
+g_{2D}(t)|\psi|\Bigg)\psi+ S(t) e^{i \theta(x,t)},\ee where $\psi$
is the macroscopic wave function, $t$ is the time and $x$ is the
transverse direction. Eq. \ref{e1} appears as an approximate model that governs the evolution of the macroscopic wave function of a
cigar-shaped one-dimensional BEC. Here, $g_{1D}(t)$ and $g_{2D}(t)$ are the cubic and quadratic nonlinearity coefficients, describes repulsive
contact interactions between atoms carrying dipole
moments \cite{quad1} and dipole-dipole attraction \cite{quad2}. The source term, $S(t)e^{i \theta(x,t)}$ with $S(t)$ and $\theta(x,t)$
as amplitude and phase, represents the coupling of BEC reservoir to a waveguide \cite{Paul2,Yan}.
$V(x,t)$ represents the complex potential, given as $V(x,t)
=\frac{F(t)x^2}{2} + i \frac{H(t)}{2}$, where $\frac{F(t)x^2}{2}$
is time dependent parabolic trapping potential, and $H(t)$ is gain or loss
term and account for the interaction of atomic and thermal cloud. The complex potential
is PT-symmetric if the real and imaginary parts of the potential must
be an even and odd functions along the longitudinal direction, respectively, i.e.
$F(t) = F(-t)$ and $H(t) = -H(-t)$. In Ref. \cite{nixon}, authors have investigated the light propagation in optical waveguides satisying PT-symmetry along the longitudinal direction. Normalizing the time and
length in Eq. (\ref{e1}) in units of  ${\omega_\perp}^{-1}$ and $\sqrt{\frac{h}{m \omega_\perp}}$
 where ${\omega_\perp}$ defines the transverse trapping
frequency, the dimensionless form of GP equation with a source
term can be expressed as
\be \label{e2} i\frac{\partial\psi}{\partial t}=-\frac{1}{2}
\frac{\partial^2\psi}{\partial x^2}+ f(t) \frac{x^2}{2}\psi
+\frac{i}{2} h(t) \psi + g_1(t) |\psi|^2\psi +g_2(t)|\psi| \psi+ s(t)
e^{i\theta(x,t)}.\ee Eq. (\ref{e2}) is linked with  $ \frac{\delta
\mathscr{L}}{\delta \psi^*} = 0$ where Lagrangian density can be
written as
\be\label{E2} \mathscr{L} = i (\psi
\psi_t^{*}- \psi_t \psi^*) + |\psi_x|^2 +  (f(t)x^2 + i h(t))
|\psi|^2 + g_1(t)|\psi|^4 + \frac{4}{3} g_2(t) |\psi|^3 + 2 s(t)
e^{i\theta(x,t)}\psi^*, \ee
where $\psi^*(x,t)$ indicates the complex
conjugate of the wave function $\psi(x,t)$.

\section{Self-similar matter waves}
 In order to obtain the matter wave solutions of Eq. (\ref{e2}), we
choose self-similar transformation as \cite{agrawal,goyal}
     \be\label{e3} \psi(x,t)  =  A(t)~~U\left[\frac{x-x_c(t)}{W(t)},\zeta(t)\right]
     e^{i \phi(x,t)},\ee
where $A(t)$, $W(t)$ and $x_c(t)$ are the dimensionless amplitude, width
and center position of the self-similar wave, respectively. The phase is chosen
as \be\label{e4} \phi(x,t) = C_1(t)\frac{x^2}{2} + C_2(t)x +
C_3(t), \ee where $C_1(t)$, $C_2(t)$ and $C_3(t)$ are the
parameter related to the phase-front curvature, the frequency
shift and the phase shift, respectively, to be determined.
Substituting Eq. (\ref{e3}) and Eq. (\ref{e4}) into Eq.
(\ref{e2}), one obtains the quadratic-cubic NLSE with drive  as
\be\label{e5} i\frac{\partial U }{\partial \zeta} +
\frac{1}{2}\frac{\partial^2 U}{\partial \chi^2} + a_1|U|^2 U + a_2
|U| U = a_3~e^{i(\theta -\phi)}, \ee where the amplitude,
similarity variable, effective propagation distance,
 guiding-center position, cubic nonlinearity, quadratic nonlinearity, source term and phase are given as
\be\label{e6} A(t)=\frac{1}{W(t)},~~
\chi(x,t)=\frac{x-x_c(t)}{W(t)},~~ \zeta(t) = \zeta_0 +
\int_{0}^{t}\frac{ds}{W^2(s)},\ee
 \be\label{e7}  x_c(t) = W(t)\Bigg(x_0 + C_{02} \int_{0}^{t}\frac{ds}{W^2(s)}\Bigg),~~ g_1 = - a_1,~~
  g_2 = - a_2~A(t), \ee,
  \be\label{e8} s(t)=  a_3~A^3(t),~~ \phi(x,t) = \frac {x^2}{2W} \frac {\partial W}{\partial t} +
 \frac{C_{02}x}{W} -\frac{C^2_{02}}{2}\int_{0}^{t}\frac{ds}{W^2(s)}, \ee
 with $C_2(0)=C_{02}$, $x_c(0)=x_0$, $W(0) = 1$, and $a_1$, $a_2$,
 $a_3$ are constants. Here, $a_1$ and $a_2$ are parameters for
 cubic and quadratic coefficient, respectively. The parameter $a_3$ modulate
 the source profile. For positive value of $a_3$, source has
 similar profile as amplitude `$A$' and for negative values of
 $a_3$, profile is inverted which can be obtained by inserting a
 extra phase difference of `$\pi$' between source and matter
 wave. The trapping potential and gain / loss are associated to
self-similar wave width as \be\label{e9} \frac{d^2 W}{d t
^2}-f(t)W = 0,~~~ h(t) = -\frac{d[ln W(t)]}{dt}. \ee

As stated earlier, transformation given by Eq. (\ref{e3}) reduces Eq. (\ref{e2})
to the constant coefficient quadratic-cubic NLSE with external source given by Eq. (\ref{e5}).
This equation is considered in the work of Pal et al. \cite{pal} to obtain a wide class of localized solutions under different parametric constraints. For all these localized solutions of Eq. (\ref{e5}), the corresponding self-similar matter wave
solutions of Eq. (\ref{e2}) can be obtained by means of the reverse
transformation variables and functions. In order to make
the paper self-contained, we sketch the essential steps of Ref. \cite{pal}.
We consider the following form of travelling wave solution for Eq. (\ref{e5}) \be\label{e12}
U(\chi,\zeta) = B(\xi)~~ e^{i(k \chi -\omega \zeta)},\ee
  where $\xi = \alpha(\chi-v \zeta)$ is the travelling coordinate with $k$,
$v$ and $\omega$ as wave parameters, and $\alpha$ is a constant.
Substituting Eq. (\ref{e12}) into Eq. (\ref{e5}), assuming
$\theta(x,t)= k \chi- \omega\zeta + \phi(x,t)$,
   and separating the
  real and imaginary parts, we obtain \be\label{e13} v = k,\ee and
   \be\label{e14} \frac{1}{2}\alpha^2 B'' + a_1 B^3 + a_2
  B^2 +\sigma B =a_3, \ee
  where $\sigma = \omega -\frac{1}{2}k^2 $. Using a scaling transformation
  \be\label{e15} B(\xi)=  \rho(\xi) +\beta, \ee
  where $\beta$ is a scaling parameter, Eq. (\ref{e14})
  reduces to \be\label{e16}\frac{1}{2} \alpha^2 \rho'' + a_1\rho^3 + \eta
  \rho^2+ \epsilon \rho -\delta = 0, \ee
where $\eta = 3a_1\beta + a_2$, $\epsilon = 3a_1 \beta^2 + 2a_2\beta + \sigma
$, $\delta = -(a_1\beta^3 + a_2 \beta^2 + \sigma\beta
-a_3)$. Here, $\eta$, $\epsilon$ and $\delta$ are effective quadratic
coefficient, effective linear term and effective source coefficient. This reduction
helps us to obtain a class of soliton solution by taking $\eta$, $\epsilon$ and $\delta$
equal to zero in Eq. (\ref{e16}) while the actual source, quadratic coefficient
and linear term is non-zero in the Eq. (\ref{e14}). We studied the evolution of matter waves for two types of
trapping potential for specific
form of width profile as per Eq. (\ref{e9}), pertaining to the
condition that complex potential should be PT-symmetric. Here, we
have choosen sech-type and Gaussian-type profiles for  trapping
potential to study the dynamics of BEC.
\subsection{Sech-type trapping potential}
 For $W(t) = a + \mbox{sech(t)}$,
the trapping potential and gain / loss functions [refer Eq.
(\ref{e9})] takes the form, \be\label{es} f(t)=
\frac{[-3+\text{cosh}(2 t)]~\text{sech}^3(t)}{2~
 [a+\text{sech}(t)]},~~ h(t)= \frac{\text{sech}(t)~\text{tanh}(t)}{a+\text{sech}(t)}.\ee

\begin{figure}[ht!]
\begin{center}
\includegraphics[scale=0.7]{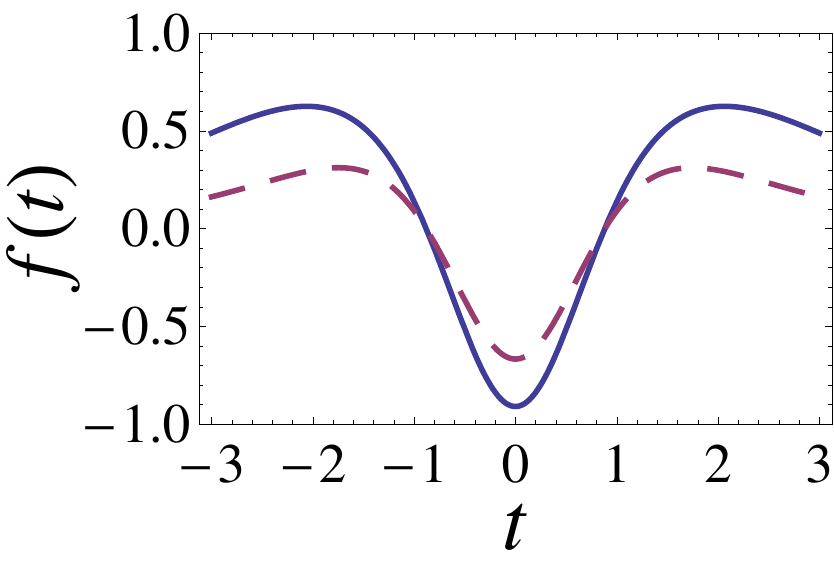}~~~ \includegraphics[scale=0.7]{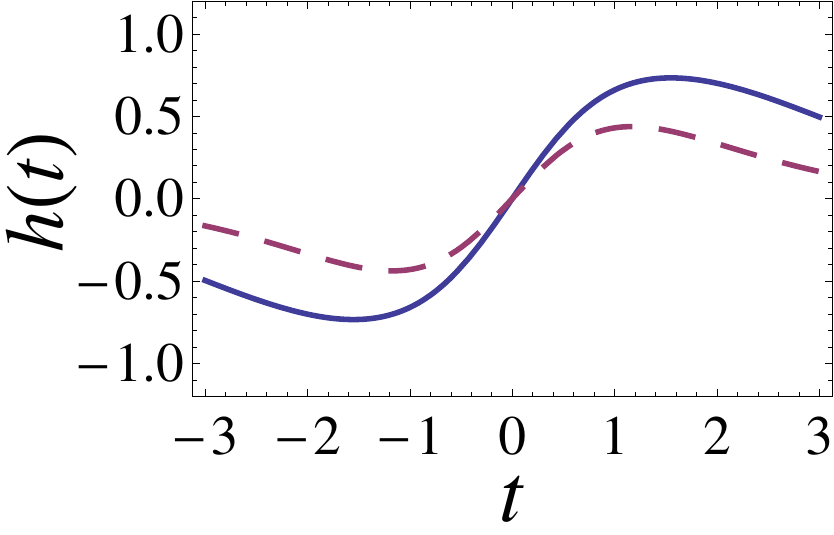}
\caption{\label{width} Profiles of trapping potential $f(t)$ for
$a = 0.1$ (solid line), $a = 0.5$ (dotted line) and gain / loss function
$h(t)$ for $a = 0.1$ (solid line), $a = 0.5$ (dotted line),
respectively.}
\end{center}
\end{figure}

The profile of trapping potential and gain / loss functions is shown
in Fig. (\ref{width}) for different values of `$a$'. As value of
`$a$' increases, the magnitude of trapping potential decreases.
From Fig. (\ref{width}), one can observe that trapping potential `$f(t)$' is
an even-function whereas `$h(t)$' is an odd-function of time
variable, which is the necessary conditions for a complex
potential to be PT-symmetric. Here, the free parameter `$a$'
should be non-zero, as $a=0$ leads to singularity in quadratic
nonlinearity. Next, we will study the evolution of self-similar
waves for this choice of trapping potential.\\

\centerline{\textbf{Bright and kink self-similar waves}} For $\eta = 0$ and $\delta = 0$,
Eq. (\ref{e16}) reduces to well known cubic elliptic equation which admits either
bright or dark solitons depending upon the sign of cubic
nonlinearity coefficient. These conditions put the constraint on
various parameters as follows \be\label{ei} \beta = \frac{-a_2}{3 a_1},
\epsilon = \sigma -\frac{a_2^2}{3
 a_1}, \sigma = \frac{2a_2^2}{9 a_1}-\frac{3 a_1 a_3}{a_2}. \ee For $a_1>0$ and $\epsilon<0$ which implies
$\sigma < \frac{a_2^2}{3 a_1}$, the cubic elliptic equation possesses bright soliton expressed as \cite{Shabat}
 \be\label{eb} \rho (x,t) =
\sqrt{\frac{-2\epsilon}{a_1}} \mbox{sech}
\Bigg(\sqrt{\frac{-2\epsilon}{\alpha^2}} \xi \Bigg).\ee The
condition $\sigma = \frac{2a_2^2}{9 a_1}-\frac{3 a_1 a_3}{a_2}$
fixes the source term as  $a_3 > \frac{-a_2^3}{27 a_1^2}$ or $a_3
< \frac{-a_2^3}{27 a_1^2}$ for $a_2$ to be positive or negative,
respectively. It implies that the model possesses bright
self-similar waves for both cases such as cubic and quadratic
nonlinearities are of same or competitive nature depending upon
the sign of $a_1$ and $a_2$ to be same or opposite. Using Eq. (\ref{eb}) along with Eqs. (\ref{e15}) and
(\ref{e12}) into
 Eq. (\ref{e3}), the complex wave solution for  Eq. (\ref{e2}) can be written as
 \be\label{e20} \psi(x,t) =\frac{1}{W(t)}\Bigg[\sqrt{\frac{-2\epsilon}{a_1}}
 \mbox{sech}
\Bigg(\sqrt{\frac{-2\epsilon}{\alpha^2}} \xi \Bigg)-
\frac{a_2}{3a_1}\Bigg]~e^{i(k\chi- \omega\zeta + \phi(x,t))}.
\ee

\begin{figure}[h!]
\begin{center}
\includegraphics[scale=0.55]{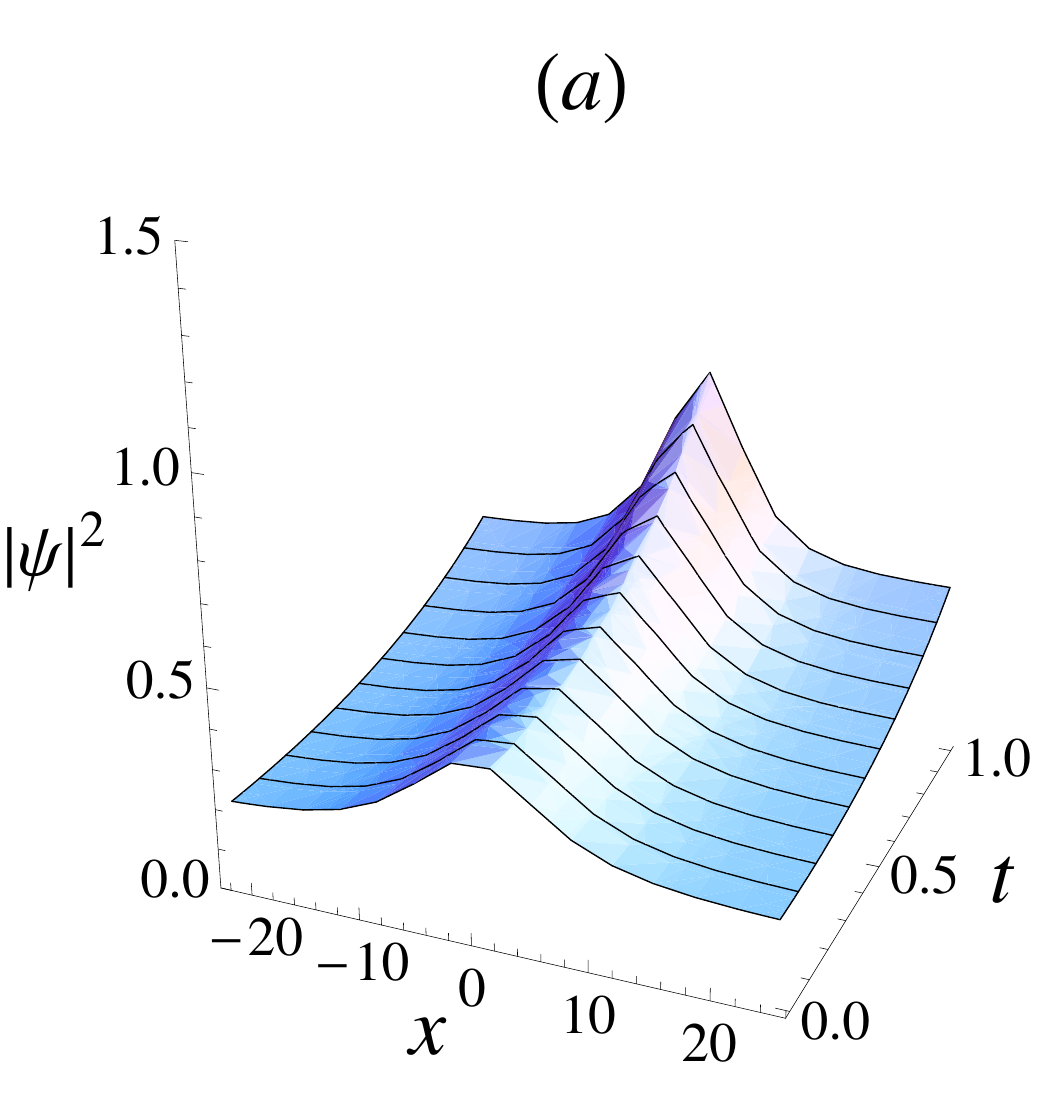}\includegraphics[scale=0.55]{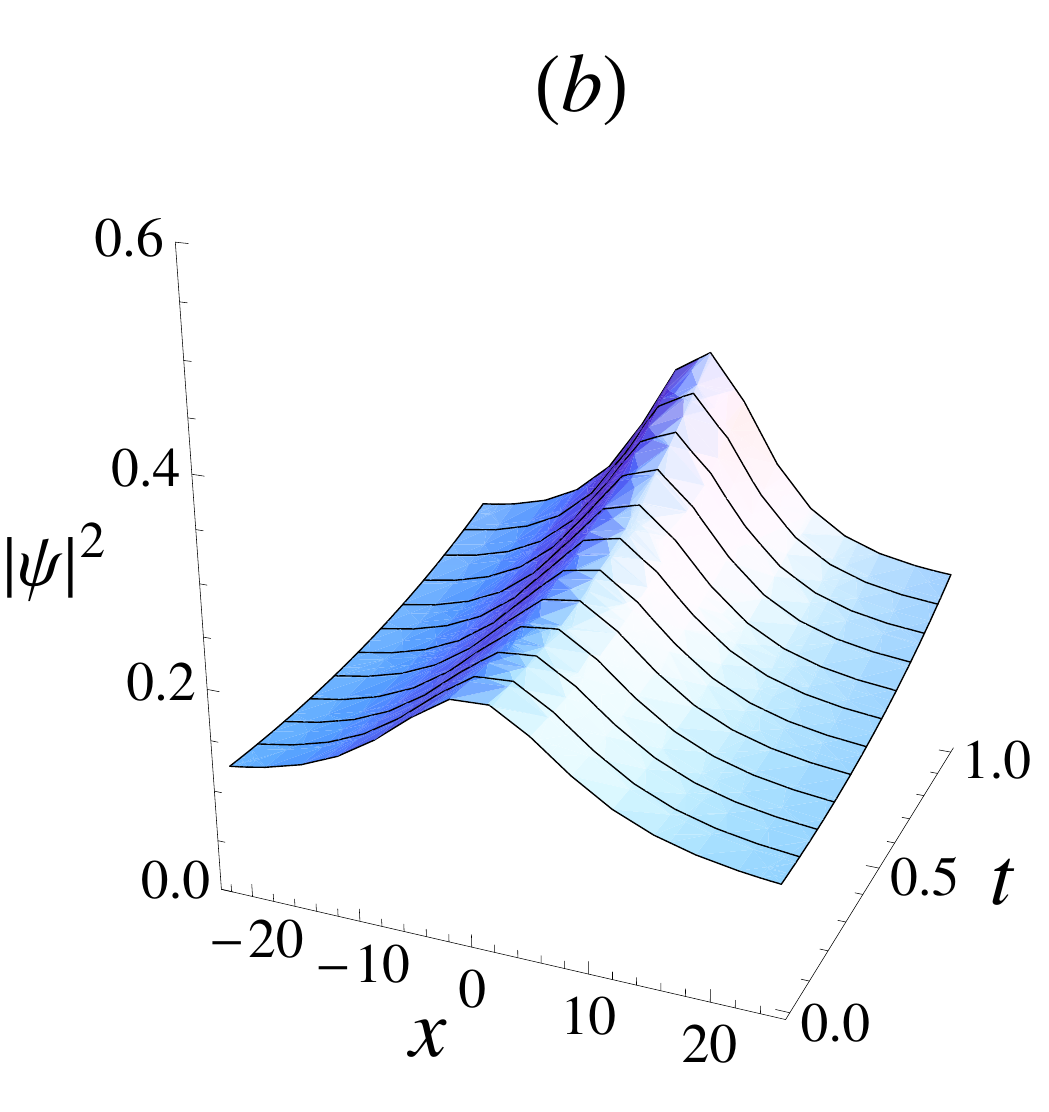}
\caption{\label{Fig1} Intensity distribution of bright self-similar matter waves for $a_1=0.8$, $a_2=-1.2$,
$a_3=0.09$ and for different values of `$a$', as (a) $a = 0.1$ and (b) $a =
0.5$, respectively. The values of other parameters used in the
plots are mentioned in the text.}
\end{center}
\end{figure}

\begin{figure}[h!]
\begin{center}
\includegraphics[scale=0.55]{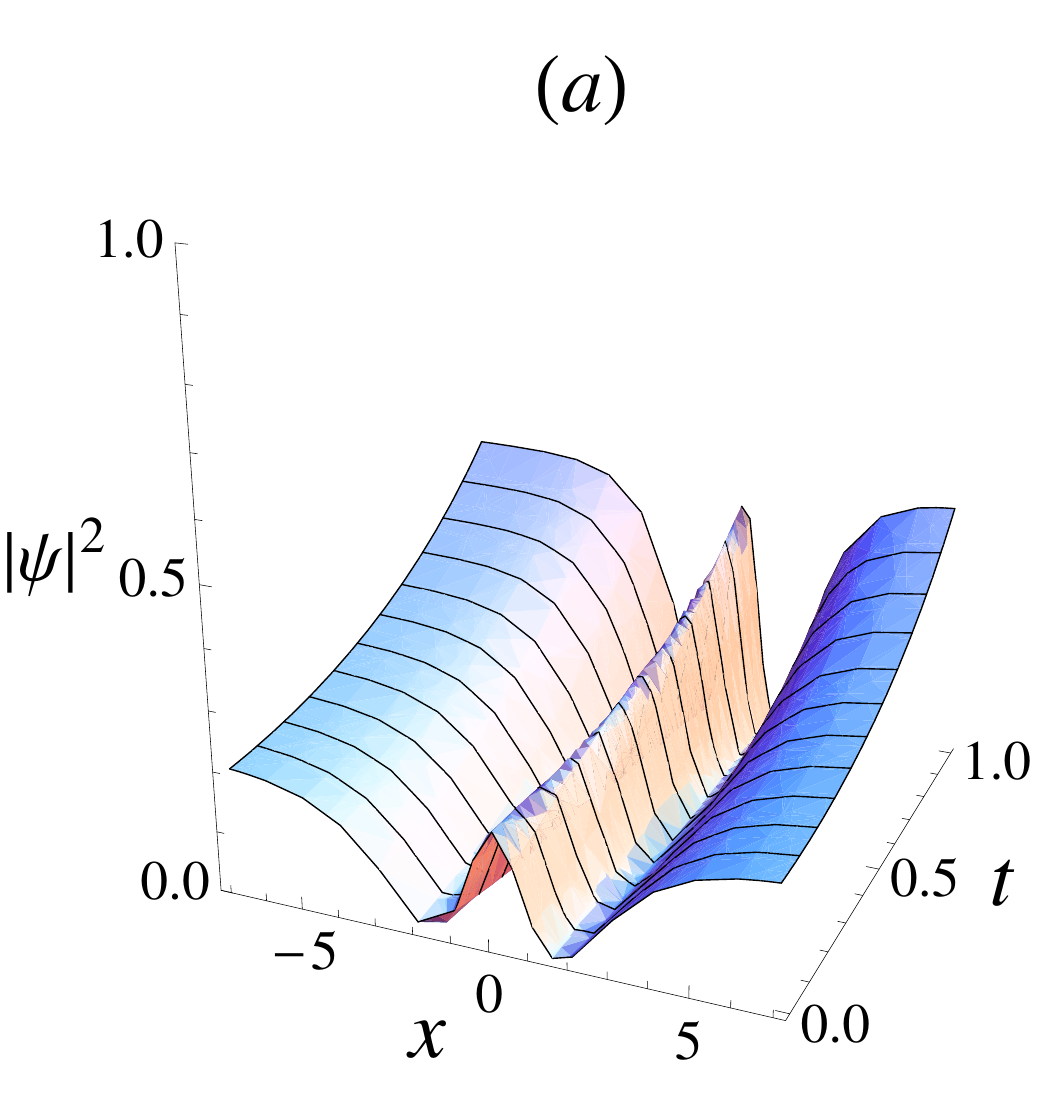}\includegraphics[scale=0.55]{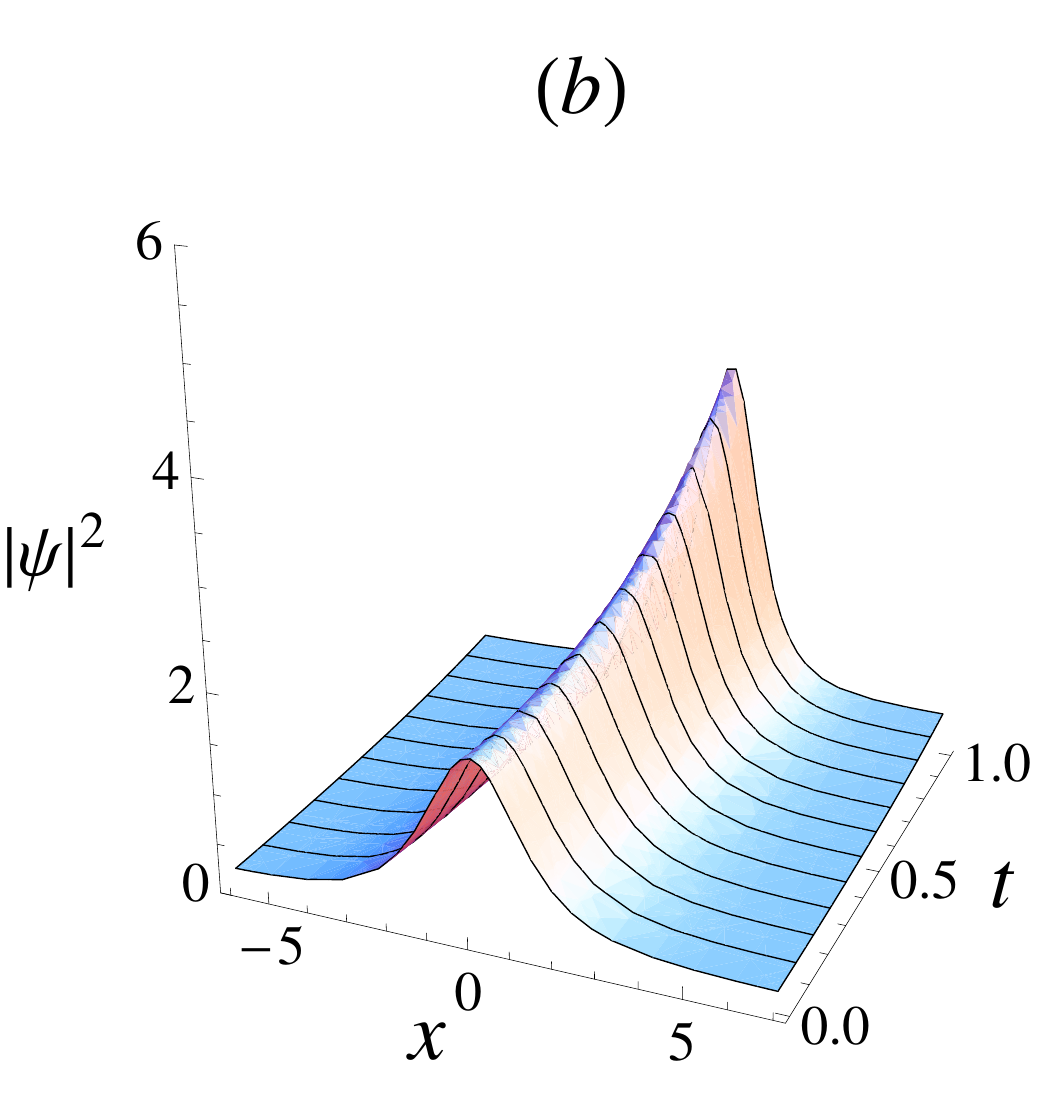}
\caption{\label{Figi} Intensity distribution of bright self-similar matter waves for $a=0.1$, $a_1=0.8$, and (a)
 $a_2 = 1.2$, $a_3 = 0.09$; (b) $a_2 = -1.2$, $a_3 = -0.09$. The other parameters chosen are same as in Fig.
 \ref{Fig1} (a). }
\end{center}
\end{figure}

We have depicted the Intensity distribution, $|\psi(x,t)|^2$, of bright self-similar waves in Fig.
 (\ref{Fig1}) for competitive nonlinearities and different values
 of `$a$' as $a=0.1$ and $a=0.5$, respectively. The nonlinearity
 coefficients are chosen as $a_1=0.8$, $a_2=-1.2$ and source
 coefficient as $a_3=0.09$. The other parameters used are $v=1$, $\alpha=1$, $\zeta _0=0$, $x_0=0$ and $C_{02}
 =0.3$. From plot, one can observe that for small values of `$a$',
 that is large amplitude of trapping potential as shown in Fig.
 (\ref{width}), the intensity of bright self-similar matter wave is more. Hence, the self-similar waves gets more
intensive and compressive as value of `$a$' decreases. Thus one can amplify
the propagating wave by judicious choice of parameter `$a$'.
Further, we have plotted the intensity profile of bright
self-similar matter wave in
 Fig. (\ref{Figi}) to analyze the effect of same sign of nonlinearities and negative magnitude of source coefficient.
 For same sign of nonlinearities chosen as $a_1 = 0.8$ and $a_2 =
1.2$, we observe W-shaped self-similar waves \cite {Triki} as
shown in Fig. \ref{Figi}(a) compared to the bell-shaped
self-similar waves for competitive nature presented by Fig. \ref{Fig1}(a). If sign of source coefficient is reversed which can
be done by adding a extra phase difference of $\pi$ in the source
profile, one can observe a very intensive and compressive
self-similar wave depicted in the Fig. \ref{Figi}(b) as compared
to Fig. \ref{Fig1}(a). Hence, the presence of source term helps to
amplify the propagating waves in a controlled manner for the
specific choice of source coefficient.

\par For $a_1 < 0$ and $\epsilon > 0$ which implies $\sigma >
\frac{a_2^2}{3 a_1}$, the cubic elliptic equation possesses dark
soliton reads as \cite{Kivshar}

\be\label{ed} \rho (x,t) = \sqrt{\frac{-\epsilon}{a_1}}
\mbox{tanh} \Bigg(\sqrt{\frac{\epsilon}{\alpha^2}} \xi \Bigg),\ee
with the same condition on source parameter as for bright soliton.
Using Eq. (\ref{ed}) along with  Eq. (\ref{e15}) and Eq.
(\ref{e12}) into
 Eq. (\ref{e3}), the complex wave solution for  Eq. (\ref{e2}) can be written as
 \be\label{e20} \psi(x,t) =\frac{1}{W(t)}\Bigg[\sqrt{\frac{-\epsilon}{a_1}}
 \mbox{tanh} \Bigg(\sqrt{\frac{\epsilon}{\alpha^2}} \xi \Bigg)-
\frac{a_2}{3a_1}\Bigg]~e^{i(k\chi- \omega\zeta + \phi(x,t))}.
\ee

\begin{figure}[h!]
\begin{center}
\includegraphics[scale=0.55]{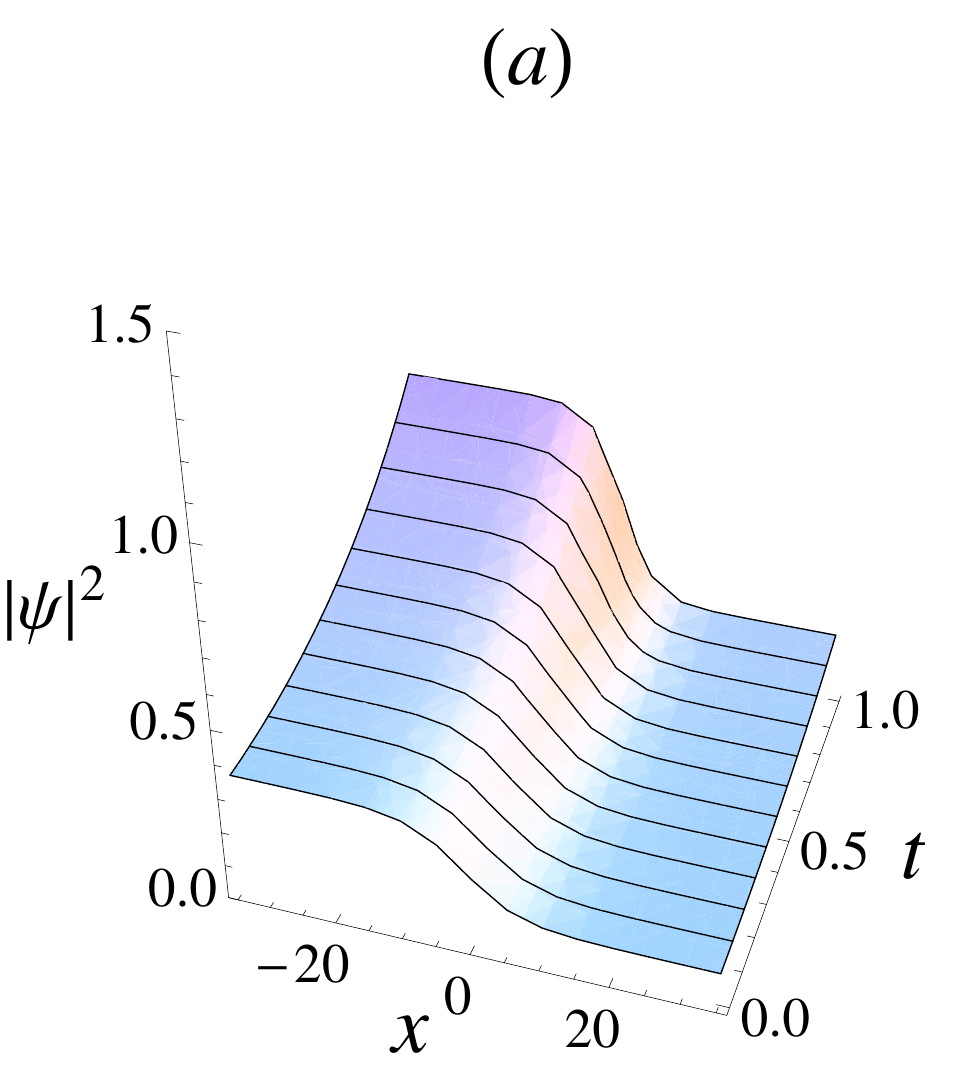}\includegraphics[scale=0.55]{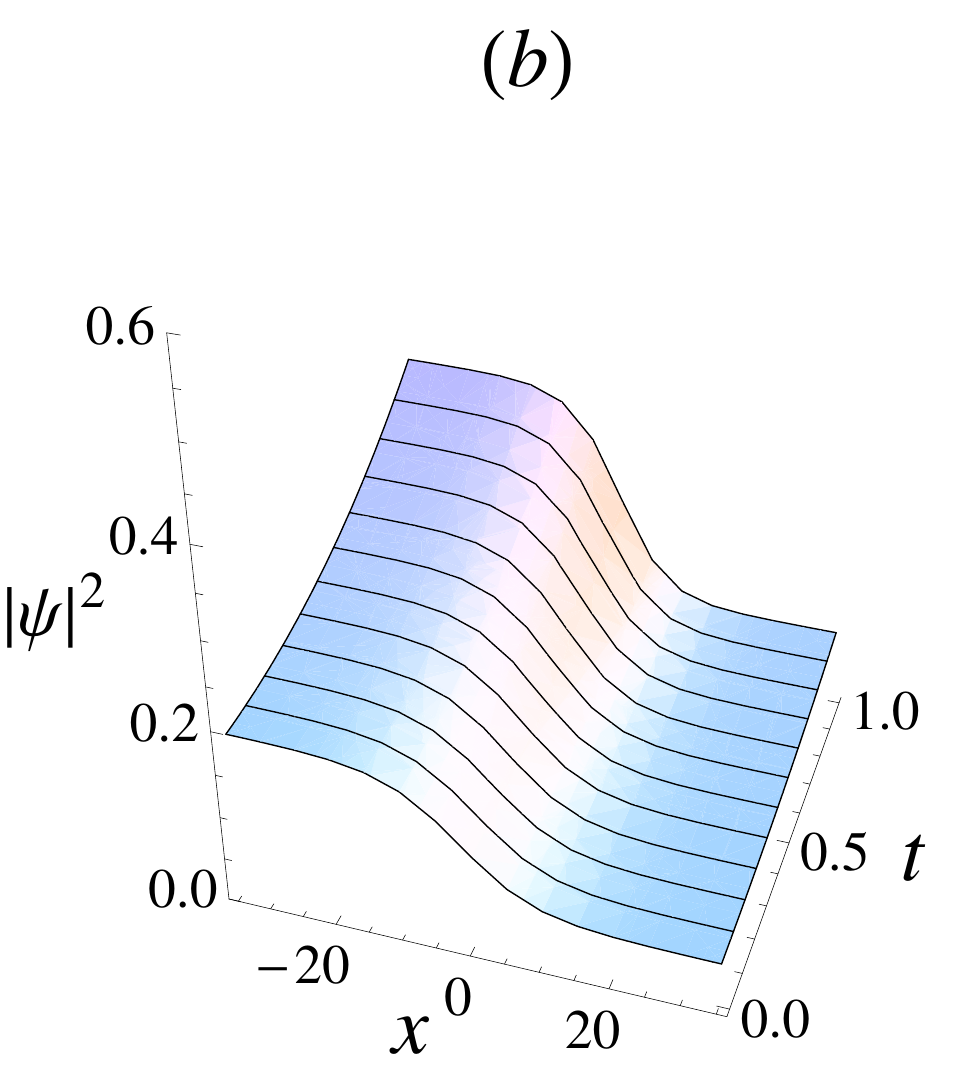}
\caption{\label{Fig2} Intensity distribution of anti-kink self-similar matter waves for $a_1=-0.8$, $a_2=-1.2$,
$a_3=0.09$ and for different values of `$a$', as (a) $a = 0.1$ and (b) $a =
0.5$, respectively. The values of other parameters used in the
plots are mentioned in the text.}
\end{center}
\end{figure}

\begin{figure}[h!]
\begin{center}
\includegraphics[scale=0.55]{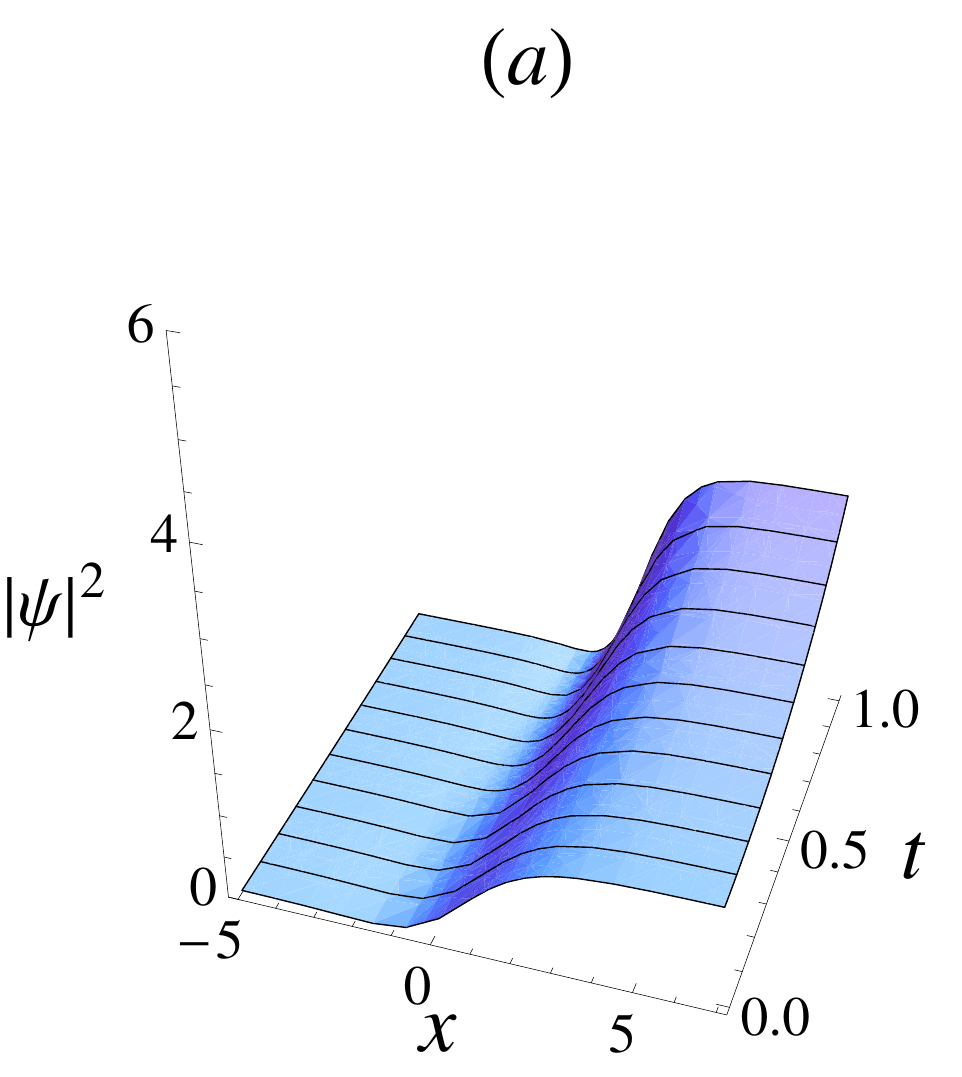}\includegraphics[scale=0.55]{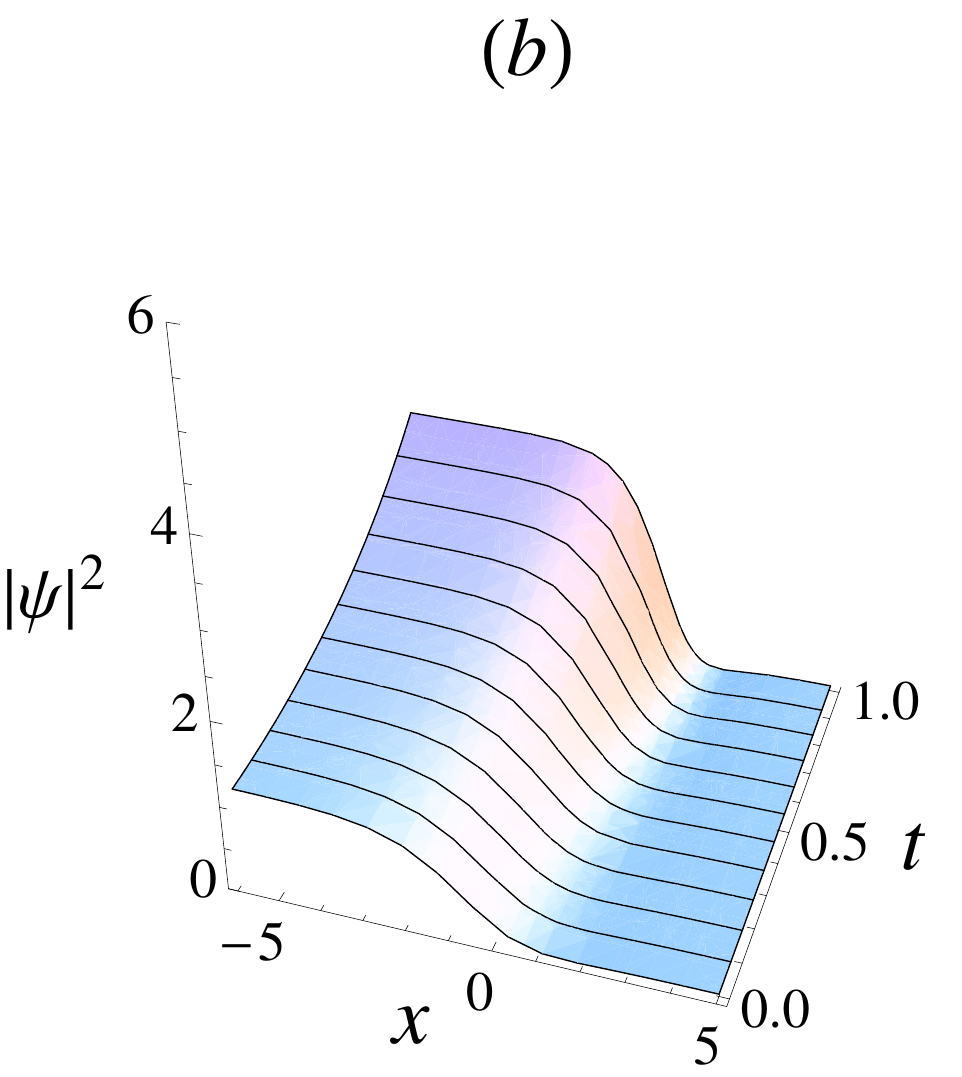}
\caption{\label{Fig3} Intensity distribution of kink and anti-kink self-similar matter waves for $a=0.1$, $a_1=-0.8$, and (a)
 $a_2 = 1.2$, $a_3 = 0.09$; (b) $a_2 = -1.2$, $a_3 = -0.09$. The other parameters chosen are same as in Fig. \ref{Fig2}(a).}
\end{center}
\end{figure}

We present the intensity distribution of these self-similar matter waves for same
sign of nonlinearities and different values of `$a$', as $a=0.1$
and $a=0.5$, respectively. As depicted in Figs. \ref{Fig2}(a) and
\ref{Fig2}(b), we have anti-kink self-similar waves which are
intensive for small values of `$a$' as compared to large values. The
other parameters are chosen as $a_1=-0.8$, $a_2=-1.2$, $a_3=0.09$,
$v=1$, $\alpha=1$, $x_0=0$, $\zeta _0=0$ and $C_{02}=0.3$. In Fig.
\ref{Fig3}(a), we have shown the intensity distribution of self-similar matter wave, given by Eq. (\ref{e20}),
for competitive nature of nonlinearities such as $a_1=-0.8$ and
$a_2=1.2$. For this case, the profile of wave is reversed and one
can observe the kink self-similar wave. Like bright self-similar matter
waves, here also intensity of waves can be increased for negative
value of source coefficient, as shown in the Fig. \ref{Fig3}(b).\\

\centerline{\textbf{Lorentzian-type self-similar waves }} For
$\epsilon = 0 $ and $\delta = 0$, Eq. (\ref{e16}) reduces to
\be\label{e25} \frac{1}{2}\alpha^2 \rho'' + a_1\rho^3 + \eta\rho^2
= 0.\ee Eq. (\ref{e25}) have Lorentzian-type algebraic soliton solutions
\cite{alka,pal2} for $a_1 > 0$, expressed as \be\label{e26}
\rho(\xi) = \frac{P}{Q+ \xi^2}, \ee where $P = \frac{-3 \alpha^2}{
\eta}$ and $ Q = \frac{9a_1 \alpha^2}{4 \eta^2} $. Subject to the
conditions $\epsilon = \delta =0 $, the parameters $\beta$ and
$\sigma$ can be obtained from equations, \be\label{e27} 2 a_1
\beta^3 + a_2 \beta^2 + a_3 =0 \ee and
 \be\label{e28} \sigma = -(3 a_1 \beta^2 + 2 a_2 \beta), \ee
respectively. Using Eq. (\ref{e26}) along with  Eqs. (\ref{e15}) and
(\ref{e12}) into
 Eq. (\ref{e3}), the complex wave solution for  Eq. (\ref{e2}) can be written as
 \be\label{e29} \psi(x,t) =\frac{1}{W(t)} \Bigg( \frac{P}{Q+ \xi^2} +\beta \Bigg)~e^{i(k\chi- \omega\zeta + \phi(x,t))}. \ee

\begin{figure}[h!]
\begin{center}
\includegraphics[scale=0.55]{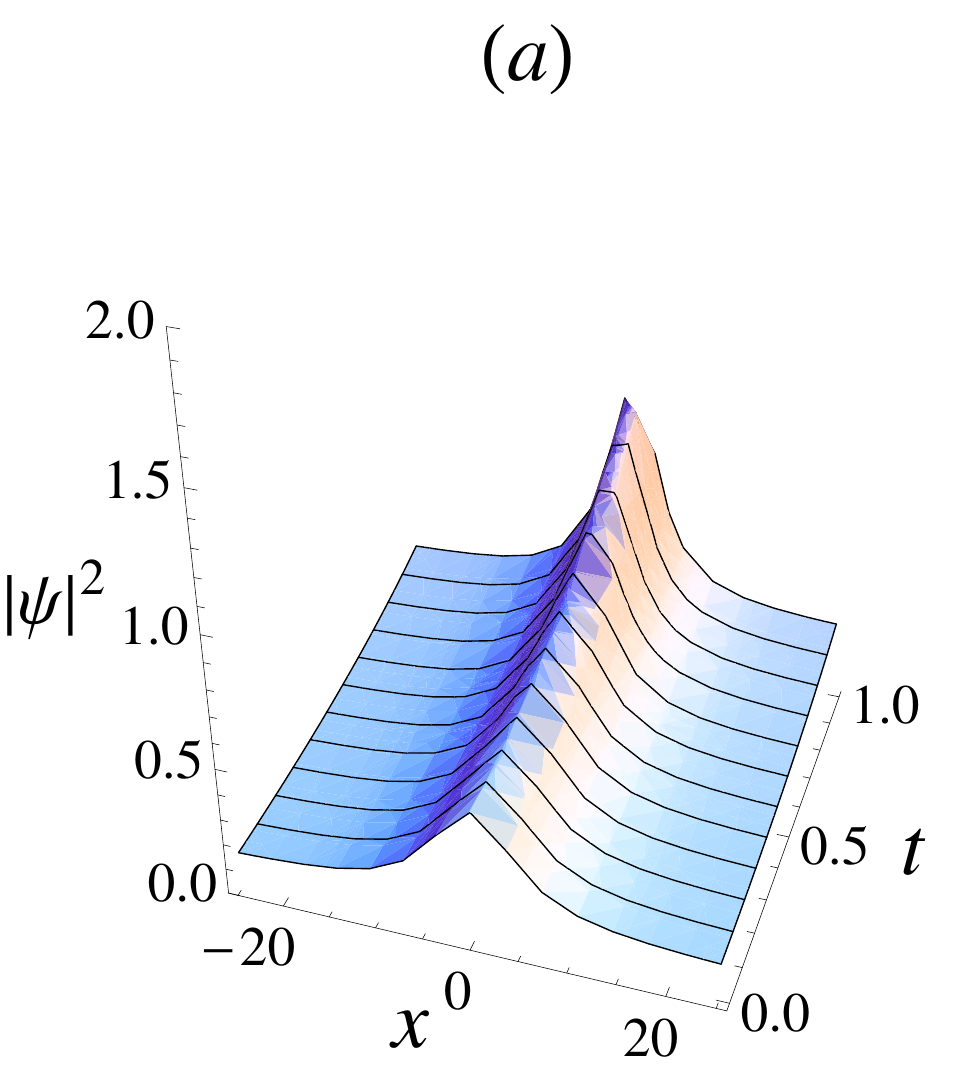}\includegraphics[scale=0.55]{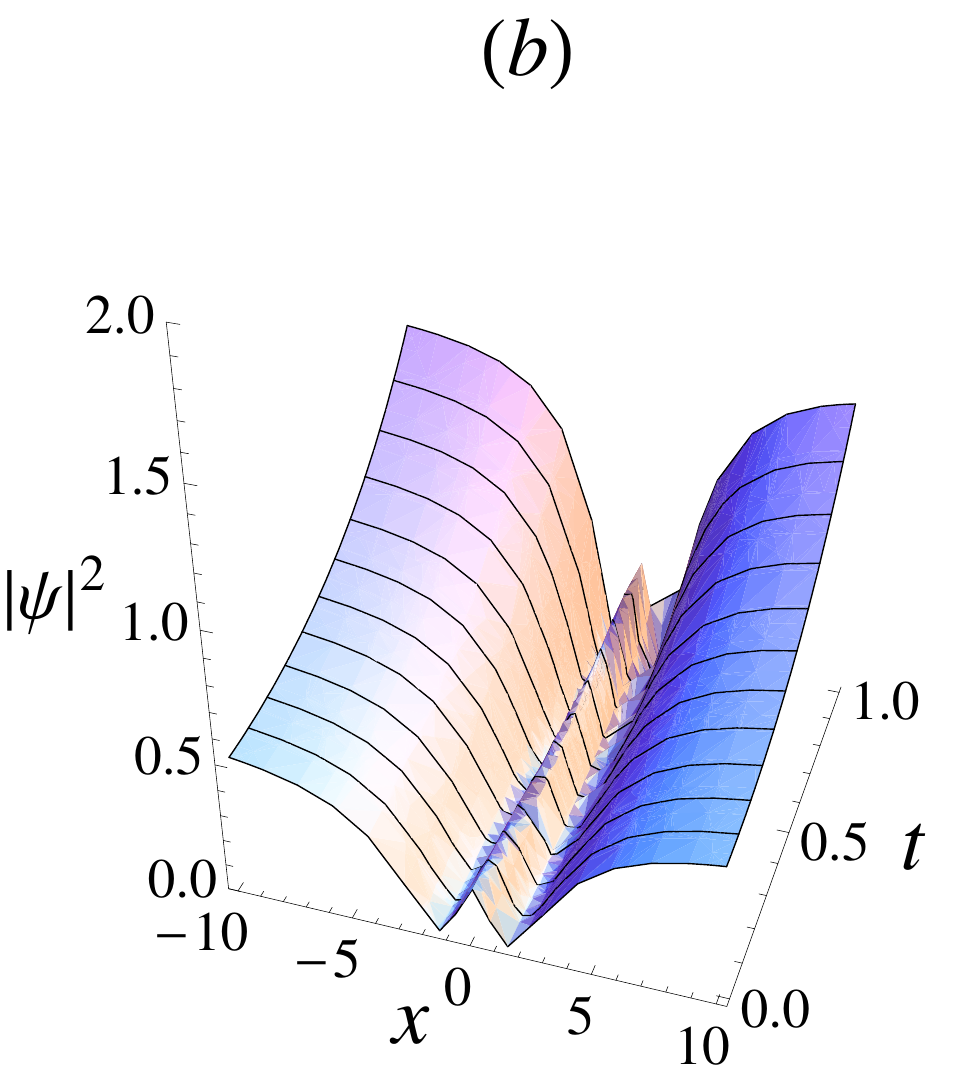}
\caption{\label{FigL} Intensity distribution of Lorentzian-type self-similar matter waves for $a=0.1$, $a_1=0.8$,
$a_3=0.09$ and for different values of $a_2$, as (a) $a_2 = -1.2$ and (b) $a_2 =
1.2$, respectively. The values of other parameters used in the
plots are mentioned in the text.}
\end{center}
\end{figure}

 The value of $\beta$ and $\sigma$ comes out from Eq. (\ref{e27})
and Eq. (\ref{e28}) by choosing values of model parameters $a_1$,
$a_2$ and $a_3$ independently. Here, the constraint condition,
$a_1>0$, is same as for bright self-similar matter wave discussed
earlier. To study the difference between these two solution, we
depicted the intensity distribution of Lorentzian-type self-similar matter waves in
Fig. \ref{FigL} (a), for same set of model parameters used in Fig.
\ref{Fig1} (a), such as competitive nonlinearities $a_1=0.8$ and $a_2=-1.2$, and other parameters as $a_3=0.09$, $a=0.1$,
$\omega=1.08$, $\alpha=1$, $\zeta _0=0$, $x_0=0$ and $C_{02}
 =0.3$. For these choices
of model parameters, the scaling parameter found to be $\beta = 0.04021$ and velocity of Lorentzian-type self-similar matter wave
is $v=1.003$, compared to $v=1$ for Fig. \ref{Fig1} (a), which further modulates the parameter $k$
in the model equation, arises in the phase relation $\theta(x,t)= k
\chi- \omega\zeta + \phi(x,t)$, as $k=v$. It means the model Eq. (\ref{e2})
can have different profiles of bright solitons, as presented in the Fig. \ref{Fig1} (a) and
Fig. \ref{FigL} (a), for small variation in the
parameter $k$ which modulates the amplitude and velocity of
self-similar matter waves. Further, the Lorentzian-type
self-similar matter waves are W-shaped for same sign of nonlinearities as shown in the Fig. \ref{FigL}
(b), for same set of model parameters as used for Fig. \ref{Figi}
(a), but having different values of $\omega$ and $v$, given by  $\omega=0.72$ and $v=0.8758$
which further modulates the parameter `$k$'. Hence, small variation
in $k$ modulates the amplitude and velocity of W-shaped
self-similar waves.\\

\begin{figure}[h!]
\begin{center}
\includegraphics[scale=0.55]{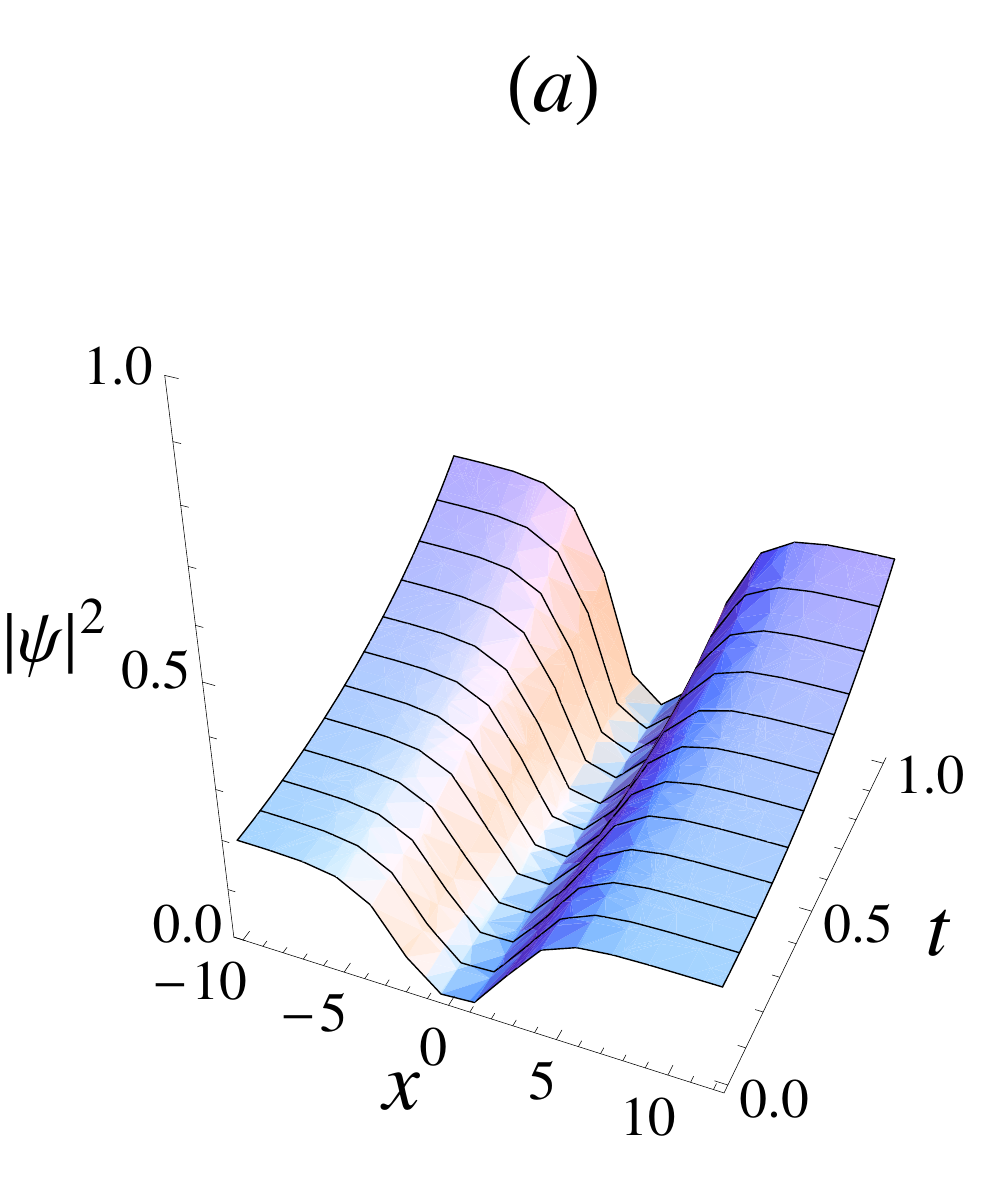}\includegraphics[scale=0.53]{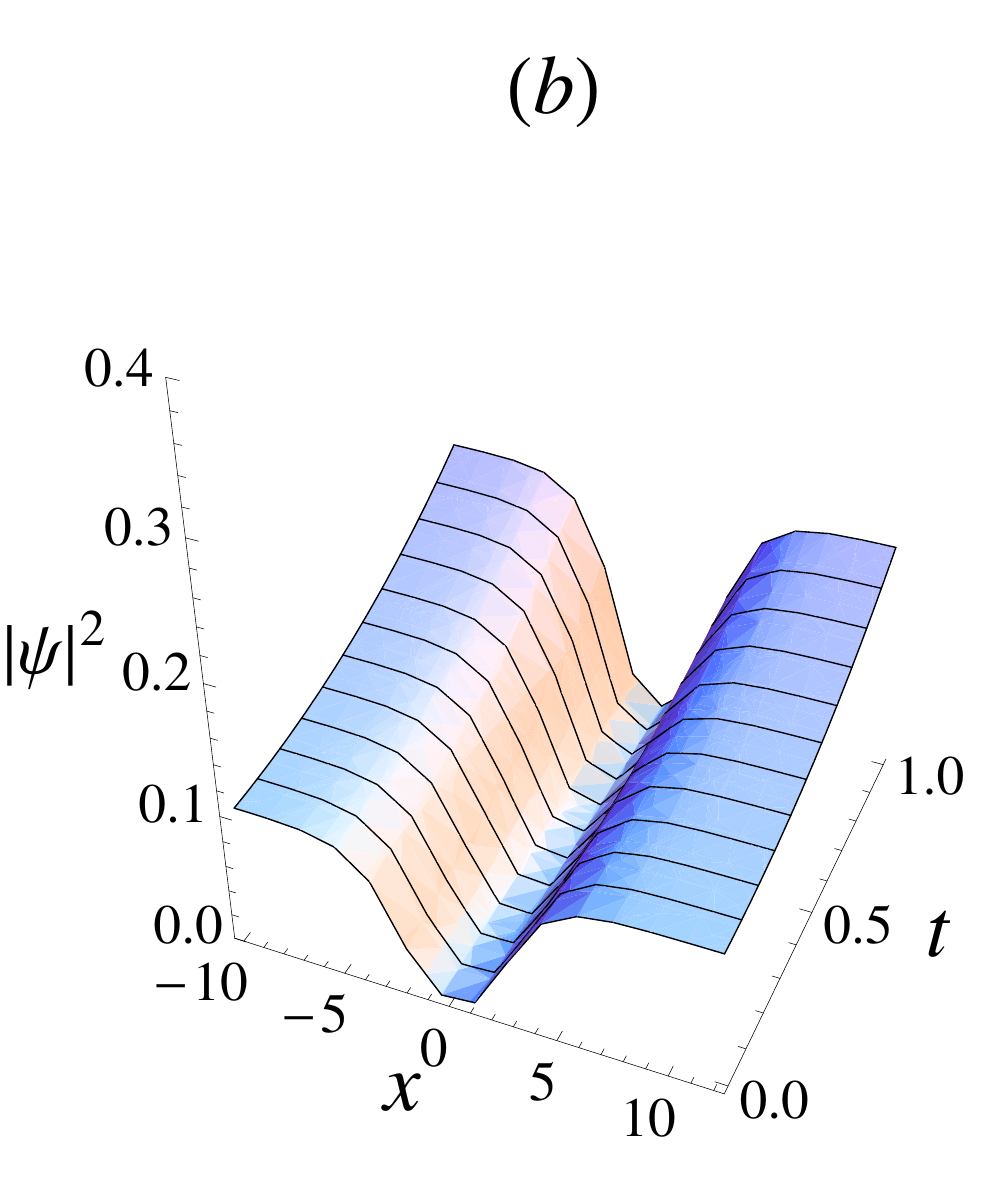}
\caption{\label{Figd} Intensity distribution of rational dark self-similar matter waves for $a_1=0$, $a_2=1.2$,
$a_3=0.09$ and for different values of $a$, as (a) $a = 0.1$ and (b) $a =
0.5$, respectively. The values of other parameters used in the
plots are mentioned in the text.}
\end{center}
\end{figure}

\centerline{\textbf{Rational dark self-similar waves}} For all the
parameters of Eq. (\ref{e16}) to be non-zero i.e. $\epsilon\neq
0$, $\eta\neq 0$, $\delta\neq 0 $ , the Eq. (\ref{e16}) have
rational dark soliton solution expressed as \cite{pal}
\be\label{erd} \rho(\xi) = \frac{p}{q + \mbox{sech}(r \xi)}, \ee
with the conditions given as
 \be\label{e23} a_1 = \frac{ q^2 r^2 \alpha^2}{p^2} (q^2-1) ,
 ~~\eta = \frac{3 q r^2 \alpha^2}{2 p} (1-2q^2), \ee
 \be\label{e24} \epsilon = \frac{(6 q^2-1)}{2} \alpha^2 r^2 , ~~\delta =  p
 q r^2 \alpha ^2. \ee
 Using Eq. (\ref{erd}) along with  Eqs. (\ref{e15}) and
(\ref{e12}) into
 Eq. (\ref{e3}), the complex wave solution for  Eq. (\ref{e2}) can be written as
 \be\label{e22}
\psi(x,t)=\frac{1}{W(t)}\Bigg[ \frac{p}{q + \mbox{sech}(r
\xi)}+\beta\Bigg]~e^{i(k\chi- \omega\zeta + \phi(x,t))}. \ee
The Eqs. (\ref{e23}) and (\ref{e24}) can be solved to obtain the
values of unknown parameters $p$, $q$ and $r$ along with one
parametric condition. Here, we have presented an interesting
case by assuming $q=1$ for which the cubic nonlinearity
coefficient `$a_1$' equals to zero and the systems induce only
quadratic nonlinearity. For $q=1$, the values of other parameters $p$
and $r$ comes out to be $\frac{5\delta}{2\epsilon}$ and
 $\pm\sqrt{\frac{2 \epsilon}{5 \alpha^2}}$, respectively along with parametric condition $\eta  = \frac{-6
 \epsilon^2}{ 25\delta}$. This condition fixes the value of $\sigma$ term
 which can be obtained from the following relation
  \be\label{es} \sigma = \frac{1}{12} \left(\beta  a_2 \pm 5 \sqrt{\beta ^2 a_2^2-24 a_2 a_3}\right).  \ee
The magnitude of the $\sigma$ estimated to be  $0.76589$ for the
parameters $a_1 = 0$, $a_2 = 1.2$ and $a_3 = -0.09$. The other parameters used are $\beta=0.5$, $v=1$, $\alpha=1$, $\zeta _0=0$, $x_0=0$ and $C_{02}
 =0.3$. The intensity
plot of rational dark self-similar matter wave in the absence of cubic
nonlinearity is shown in  Fig. (\ref{Figd}) for different value of parameter
`$a$'. As depicted from
Figs. \ref{Figd} (a) and \ref{Figd} (b), the self-similar matter wave
gets more intensive for small values of `$a$'.

\subsection{Gaussian-type trapping potential} For $ W(t) = a + \text{exp}\left(-t^2\right)$, the
trapping potential and gain / loss function [refer Eq. (\ref{e9})]
takes the form \be\label{eg} f(t) = \frac{2(2 t^2-1)}{1+a~e^{t^2}},
h(t)= \frac{2 t}{1+a~e^{t^2}}. \ee

\begin{figure}[h!]
\begin{center}
\includegraphics[scale=0.65]{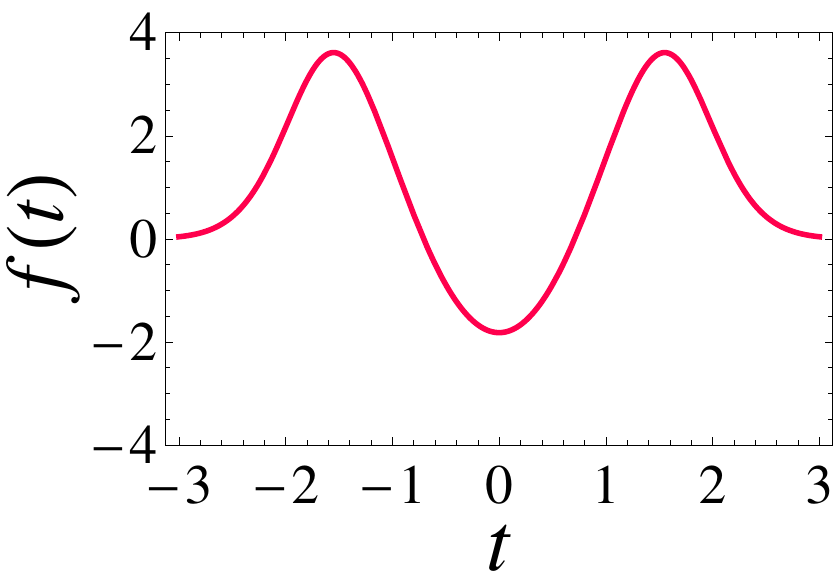}~~\includegraphics[scale=0.65]{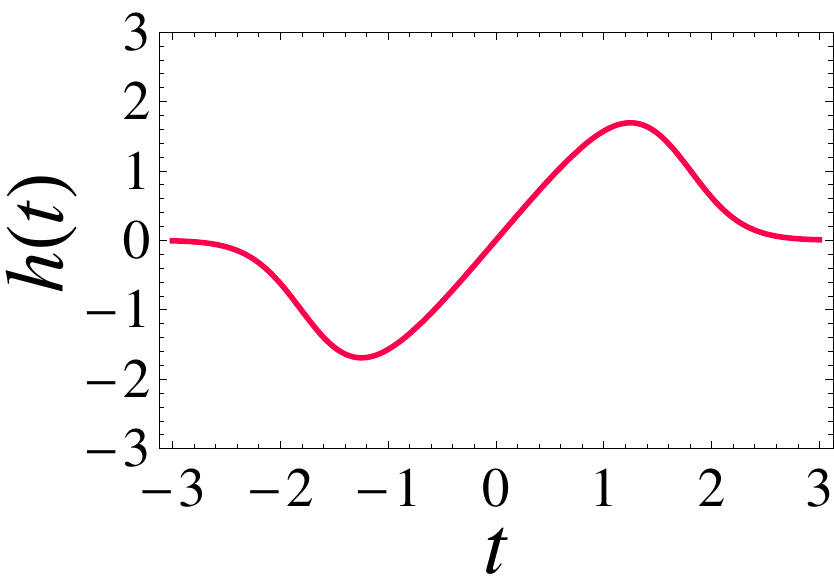}
\caption{\label{width1} Profiles of trapping potential $f(t)$ and
gain / loss function $h(t)$ for $a=0.1$.}
\end{center}
\end{figure}

\begin{figure}[h!]
\begin{center}
\includegraphics[scale=0.55]{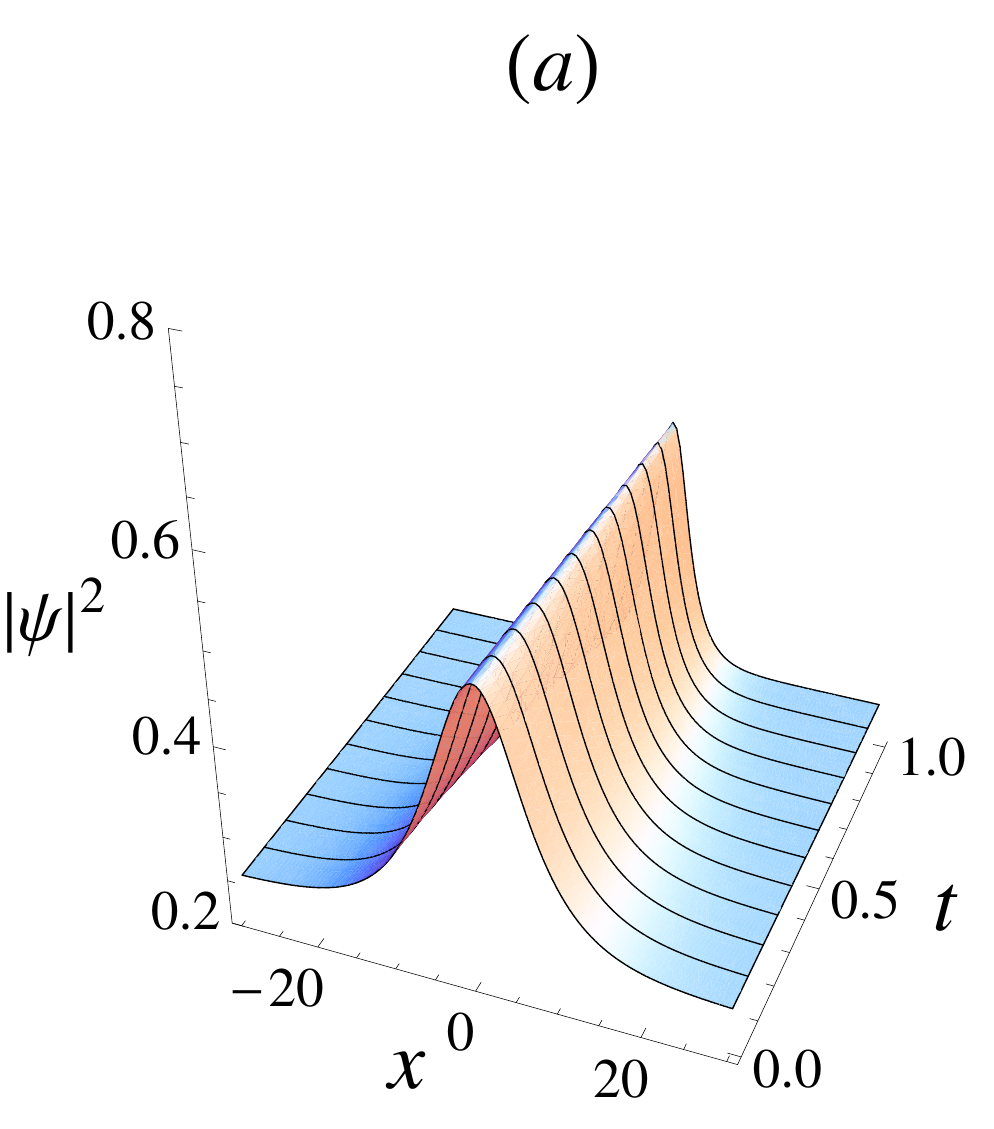}~~\includegraphics[scale=0.58]{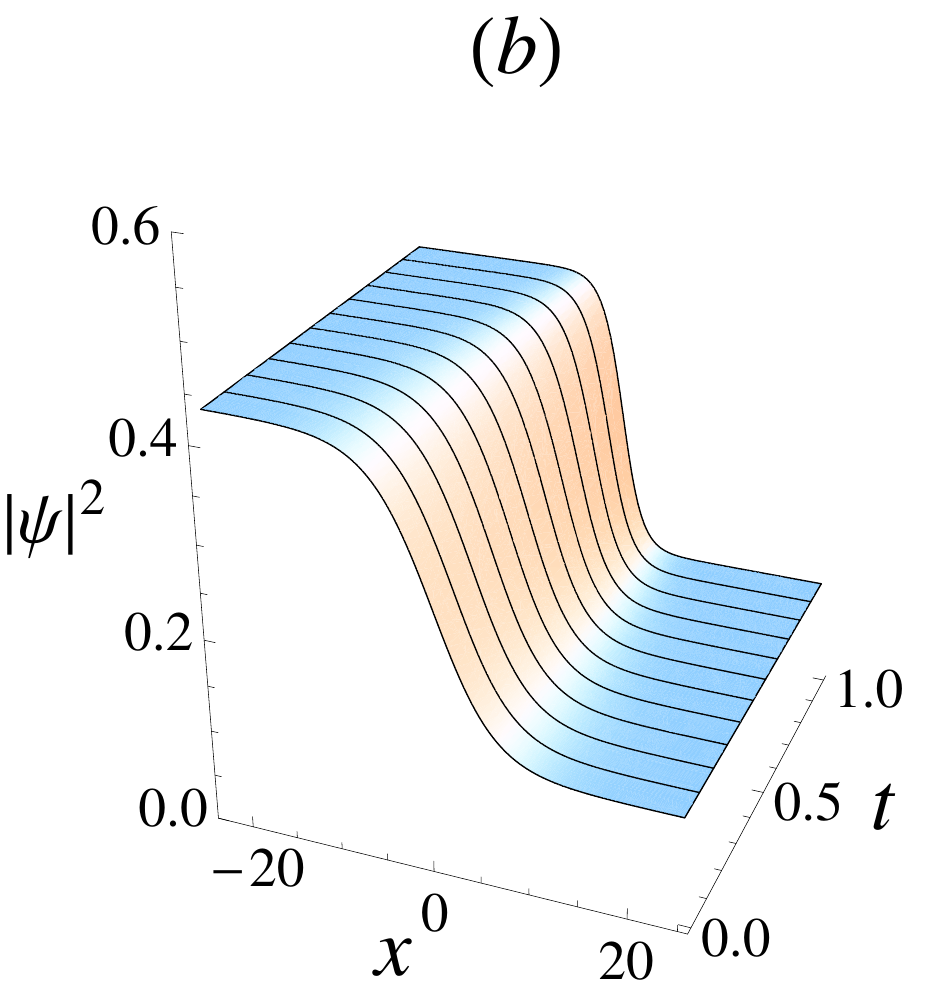}
\caption{\label{Figb} Intensity distribution of bright and kink self-similar matter waves for $a=0.1$, $a_3=0.09$
and for different nature of nonlinearities, as (a) $a_1= 0.8$, $a_2=-1.2$ and (b) $a_1=-0.8$, $a_2=-1.2$, respectively. The values of other parameters used in the plots are mentioned in the text.}
\end{center}
\end{figure}

The profile of trapping potential and gain / loss function is
presented in the Fig. (\ref{width1}), for $a=0.1$. It is clear that
$f(t)$ is an even function and $h(t)$ is an odd function of time
variable. For $a=0$, the quadratic nonlinearity turns out to be singular, so the parameter `$a$' can take
only non-zero values. But for this
choice, it is not possible to solve variables $\zeta(t)$, $x_c(t)$
and $\phi(x,t)$ explicitly given by Eqs. (\ref{e6}) - (\ref{e8}),
due to the presence of integral of reciprocal of width function. We
solve the integral implicitly using numerical method to study the evolution of bright and kink self-similar waves for this choice of
trapping potential. In  Fig. (\ref{Figb}), we depicted the
intensity distribution of bright and dark self-similar matter waves for
$a_1= 0.8$, $a_2=-1.2$ and $a_1=-0.8$, $a_2=-1.2$, respectively. The other parameters used are
$a=0.1$, $a_3=0.09$, $\alpha=1$, $v=1$, $x_0=0$, $\zeta _0=0$ and $C_{02}=0.3$.
One can study the effect of parameter `$a$' on the intensity of these waves and also observe the evolution of other self-similar waves for this
trapping potential.

\section{Conclusion}
In conclusion, we have investigated the self-similar matter waves
for driven GP equation with PT-symmetric potential in the presence
of quadratic-cubic nonlinearity. Self-similarity transformation
technique is employed to obtain bright, dark, Lorentzian-type and
kink solitons under certain
parametric constraints. The evolution of self-similar matter waves has
been depicted for sech-type and Gaussian-type trapping potentials.
We also have studied the effect of nature of nonlinearities, amplitude of trapping potential and
source profile on the intensity of self-similar matter waves. Intensity can be
made larger for specific choice of these parameters, resulting into generation of highly
energetic self-similar waves in BEC.

\section{Acknowledgment} S.P. would like to thank DST Inspire,
India, for financial support through Junior Research Fellow
[IF170725]. A.G. gratefully acknowledges Science and Engineering
Research Board (SERB), Department of Science and Technology,
Government of India for the award of SERB Start-Up Research Grant
(Young Scientists) (Sanction No: YSS/2015/001803). H.K. is thankful to
SERB-DST, India for the award of fellowship during the tenure of this work. We would like
to thank Department of Physics, Panjab University for the research
facilities.

\end{document}